\documentstyle[epsfig]{aipproc}

\begin{document}
\title{Is Nova Sco 1994 (GRO 1655-40)\\ a Relic of a GRB ?}

\author{G.E. Brown$^\star$, C.-H. Lee$^\star$, H.K. Lee$^\dagger$
 and H.A. Bethe$^{\dagger\dagger}$}
\address{$^\star$Department of Physics \& Astronomy,
        SUNY, Stony Brook, NY 11794, USA\\
  $^\dagger$Department of Physics, Hanyang University, Seoul 133-791, Korea\\
  $^{\dagger\dagger}$Floyd R. Newman Laboratory of Nuclear Studies,
       Cornell University, Ithaca, NY 14853, USA}

\def\msun{M_\odot}
\newcommand{\gsim}{\mathrel{\hbox{\rlap{\lower.55ex \hbox {$\sim$}}
                   \kern-.3em \raise.4ex \hbox{$>$}}}}
\newcommand{\lsim}{\mathrel{\hbox{\rlap{\lower.55ex \hbox {$\sim$}}
                   \kern-.3em \raise.4ex \hbox{$<$}}}}
\newcommand{\be}{\begin{eqnarray}}
\newcommand{\ee}{\end{eqnarray}}
\newcommand{\ba}{\begin{eqnarray}}
\newcommand{\ea}{\end{eqnarray}}

\maketitle

\begin{abstract}
We suggest Nova Sco 1994 (GRO 1655-40) as a possible relic of a
Gamma Ray Burster (GRB) and
Type Ib supernova (SN) explosion, showing that there is evidence both
that the black hole was spun up by accretion and that there was a supernova
explosion. We use the disc energy delivered from the rotational energy
of the black hole to power the SN, and give arguments that roughly
equal energy goes into the GRB and into the accretion
disc to power the supernova.
\end{abstract}


\section*{Introduction}

In this note we consider only the long term GRBs, of duration from
several seconds up to several minutes. This is just the dynamical
time of a He star, which we consider as progenitor.

The formation of black holes in single stars of ZAMS masses
$20-35\msun$ was proposed by Brown, Lee, \& Bethe (1999). However,
these do not lose their envelopes except in binaries. This latter
case has been studied by Brown et al. (1999) who evolve the
transient sources in this way. These have mostly low mass main
sequence companions, although in two cases the companions are
subgiants. In many more cases, binaries of $\sim 7\msun$ black
holes with companions up to nearly equal mass of the ZAMS mass of
the black hole progenitors are predicted. These
companions do not, however, fill their Roche Lobes, and
consequently are not observed. None the less, the Wolf-Rayet
progenitors of the $\sim 7\msun$ black holes  in
these binaries offer a set of progenitors for GRB's. They are
already somewhat in rotation because of the companion star. Note
that in the Brown, Lee, \& Bethe (1999) scenario, their envelopes are
removed only following He core burning, in the supergiant stage,
so there is only a short time left in their evolution for them to
lose He by wind. Thus, a substantial amount of He should be left
in the W.-R., although  most of it will have been burned to carbon
and oxygen (Woosley, Langer, \& Weaver 1993),  and the explosion
we describe below would be Type Ib.

The conclusion of the Bethe \& Brown (1999) paper was that in
order for Wolf-Rayets followed by high-mass black holes like that
in Cyg X-1 to be formed in single stars, a ZAMS mass of $\gsim
80\msun$ was necessary. This was based on the calculation of
Woosley, Langer, \& Weaver (1993) who used a too large  mass loss
rate for the He winds. These stars have been reevolved by
Wellstein \& Langer (1999) with lower mass loss rates, but the
evolution has not been carried beyond the carbon-oxygen core
stage, so we do not yet know how much the lower winds will
decrease the mass limit for evolving into a high-mass black hole.
The carbon oxygen cores still have about $33\%$ central carbon
abundances, so they clearly will not skip the convective carbon
burning stage and therefore may well end up as low mass compact
objects. These high mass Wolf-Rayet stars are the progenitors of
GRBs in Woosley's Collapsar model (MacFadyen \& Woosley 1999).

In addition to these and the high-mass black holes in the
transient sources, the coalescence of the low-mass black holes in
the Bethe \& Brown (1998) scenario of compact binary evolution
with the companion He star (Fryer \& Woosley 1996) offers another
type of generator for the long term GRBs.

All three of the above possibilities involve a He star being accreted
by a black hole. In this process the black hole will be spun up.
The energy can be extracted by the Blandford-Znajek (BZ) 1977 process,
as in Lee, Wijers, \& Brown
(1999).  The BZ power supplied into the disc will halt the inflow, and
later propel the matter outwards (Brown, Lee, Lee, \& Bethe 2000)
in a Type Ib supernova explosion.

\section*{Energetics of GRBs}

The maximum energy that can be extracted from the BZ mechanism
(Lee, Wijers, \& Brown 1999) is $ E_{max} = 0.09 M_{\rm BH} c^2$.
For a $7\msun$ black hole, such as is found in Nova Sco 1994,
$ E_{max}\simeq 1.1\times 10^{54} {\rm ergs}$.
The black hole is first formed with a mass of
at least $1.5\msun$ (Brown \& Bethe 1994). The maximum energy,
after these corrections, is still an order of magnitude greater
than the $3\times 10^{52}$ ergs used by Iwamoto et al. in the
supernova explosion. Presumably the explosion will take place
before the BZ mechanism can deliver full energy, leaving the black
hole with substantial spin energy.

Without beaming, the estimate of the energy in the jet of the GRB
(Anderson et al. 1999) is $E_{990123}=4.5\times 10^{54} {\rm ergs}$.
The BZ scenario entails substantial beaming, so this energy should
be multiplied by $d\Omega/4\pi$, which may be a small factor $\sim 0.01$.
The BZ power can be delivered at a maximum rate of
   \be
   P_{\rm BZ} = 6.7\times 10^{50}
   \left(\frac{B}{10^{15}{\rm G}}\right)^2
   \left(\frac{M_{\rm BH}}{\msun}\right)^2 {\rm erg\; s^{-1}}.
   \ee

\section*{Is Nova Sco 1994 a Relic of a GRB ?}

Several characteristics of Nova Sco 1994 (GRO 1655-40) can be understood
if it is a relic of a GRB. First of all the high space velocity
$-150\pm19$ km/s can be understood if a supernova explosion is
associated with black hole formation (Brandt et al. 1995,
Nelemans et al. 1999).

In our scenario,
first a GRB is initiated by the BZ-mechanism, following which
a Type Ib supernova explosion is begun by the energy deposited
in the fat accretion disc.
Following Brandt et al. (1995) we note that a binary symmetric in the
frame of the exploding star will be asymmetric in the center of mass
of the binary. The amount of mass that can be ejected
is constrained by the fact that if more than half of the total initial
mass is ejected, the system will become unbound. Brandt et al. consider
collapse to a $4\msun$ black hole; then at the time of collapse the
collapsing He star must have a mass of $\sim 9\msun$ or greater.

In fact, the initial black hole needs to be no more than $\sim 1.5\msun$
according to Brown \& Bethe (1994), but one would expect a substantial
amount of the carbon-oxygen core to collapse with the Fe, and also
substantial fallback of the original carbon-oxygen core, probably
burned to Fe in the explosion. In order to obtain an $\sim 7\msun$
black hole, the initial He envelope would have to be $\sim 15\msun$
corresponding to a ZAMS mass of $35-40\msun$.

Israelian et al. (1999) find a large overabundance of oxygen,
magnesium, silicon and sulpher in the F3-F8 IV/III companion star
of $1.6-3.1\msun$ orbiting the companion. These are just the
elements copiously produced in a Type Ib supernova explosion.
Contrary to Iwamoto et al. (1998), who need $0.7\msun$ of
$^{56}Fe$ to reproduce the brightness of SN 1998bw, Israelian find
no enhancement in the Fe.  In our scenario we expect the jet of
the GRB preceding the supernova explosion to go along the rotation
axis of the black hole, and the supernova explosion to be
initiated perpendicular to this in the accretion disc. The highly
nonequilibrium processes in the jet would not initially affect the
supernova, but might be expected to excite the He lines in later
stages where the expanding supernova interacts with the jet.

There are indications that the black hole in Nova Sco 1994 is spinning rapidly
(Sobczak et al. 1999, Gruzinov 1999).
We would expect the BZ central engine to stop delivering energy
following the supernova explosion which disrupts the magnetic fields.

In Appendix C of Brown, Lee, Wijers, \& Bethe (1999) the observed
GRB rate at the present time is estimated to be $\sim 0.1 $ GEM
(Galactic Event per Mega year). 
With a factor of 100 for beaming, this would require 10 GEM. 
Brown, Lee, \& Bethe (1999)
estimate the birth rate for visible transient sources in the
Galaxy to be 8.8 GeM. However, including the high-mass black-hole
binaries with companions which have not evolved to their Roche
Lobes, and therefore would not be visible, they arrive at a 25
times higher number; namely 220 GEM. (Inclusion of the ``silent"
binaries effectively removes the $q$, the ratio of masses, in the
calculation. The more massive companions are all inside of their
Roche Lobes, except for the two subgiants in the systems V404 Cyg
and XN Sco94.) 
Also, there are the other possible GRB
progenitors, Collapsars and coalescing black hole and He star,
mentioned above. Thus, there must be other severe criteria for
GRBs; e.g., high magnetic fields in the rotating He envelopes,
etc.

In evolution of the transient sources
the hydrogen envelope of
the massive star must be taken off only following He core burning, if
collapse into a high-mass black hole is to be obtained 
(Brown, Lee, \& Bethe  1999). 
In this rather
last stage the companion star would try to spin the hydrogen
envelope up towards corotation in the binary before expelling it.
With the large viscosity
from magnetic turbulance assumed in the Spruit \& Phinney (1998) argument,
the He core would be carried along, but probably in differential
rotation because the common envelope time is very short, $\sim 1$ year.

Similar considerations follow for the Fryer \& Woosley (1998)
model of coalescence of black hole with companion He star. The
common envelope evolution here need not happen as late as in the
transient source scenario, but on the other hand the predicted
merger rate is high, 380 GEM (Brown, Lee, Wijers, \& Bethe 1999).

\section*{Discussion}

We suggested Nova Sco 1994 (GRO 1655-40) as a relic of a GRB and
Type Ib SN explosion.

Our model begins from a black hole in a He star. It is assumed
that in the hypercritical accretion, the disc has a high magnetic
field. The black hole is spun up and the GRB, powered by the
Blandford-Znajek mechanism, is driven along the nearly matter-free
axis of rotation of the black hole. With high viscosity such as
follows from magnetic turbulence, the swallowing of the He star by
the black hole takes a dynamical time of the star.
The BZ mechanism stops once the accretion disc disappears and the
magnetic field disperses, leaving the black hole with some spin
energy.

We point out that power roughly equal to that driving the GRB will
be delivered into the disc made up out of hyperaccreting helium.
The delivered energy first brings the accreting matter to rest,
then drives it backwards through the accretion disc (and to the
sides of it) in a Type Ib supernova explosion. From consideration
of energetics, both the GRB and SN explosion can be powered by
several times $10^{53}$ ergs, if we take the black hole to have
mass $\sim 7\msun$ typical of those in the transient sources, but
the GRB may well take off and the explosion begin before maximum
energy is delivered.

In a more detailed paper (Brown, Lee, Lee, \& Bethe, 2000), 
we show that
hypercritical accretion onto a black hole in the middle of a
self-similar accretion disc of Narayan \& Yi (1994) type will spin
the black hole up to $\gsim 90\%$ of maximum, so that $\sim
10^{54}$ ergs is available. Running the plasma current through a
current (Thorne 1986) around the black hole and through the
accretion disc, we argue that roughly equal fraction of this
energy are available to power the GRB and the Type Ib supernova,
the former by energy delivered in the loading area up the rotation
axis of of the black hole and the latter in the accretion disc. In
a more detailed paper (Lee, Brown, \& Wijers 1999) we show that
how the plasma can pass through magnetosonic points, etc.

\section*{ACKNOWLEDGMENTS}

We would like to thank  Ralph Wijers and
Stan Woosley for useful discussions.
This work is supported partially
by the U.S. Department of Energy Grant No. DE-FG02-88ER40388.
HKL is supported also in part  by
KOSEF Grant No. 985-0200-001-2 and  BSRI 98-2441.

\end{document}